\documentclass[epj]{svjour}
\usepackage{import}
\usepackage{lineno,hyperref}
\usepackage{amsmath}
\usepackage{graphicx}
\usepackage{xcolor}
\usepackage{booktabs,array,dcolumn}
\usepackage{amssymb}
\usepackage[utf8]{inputenc}
\usepackage{cite}

\newcommand {\lamu}{\lambda_{\text{u}}}
\newcommand {\E}  {\varepsilon}
\newcommand {\om} {\omega} 
\newcommand {\Om} {\Omega}

\newcommand {\Nacc} {N_{\text{acc}}}

\newcommand {\Lcr}  {L_{\text{cr}} }
\newcommand {\Lp}   {L_{\text{p}}}
\newcommand {\Lch}   {L_{\text{ch}}}
\newcommand {\Ltot} {L_{\text{tot}}}

\newcommand {\calA} {{\cal A}}

\newcommand {\meanL} {{\langle L \rangle }}
\newcommand {\meanN} {{\langle N \rangle }}

\newcommand {\ach}  {a_{\text{ch}}}

\begin{document}
\title{Channeling of electrons and positrons in straight 
        and periodically bent diamond(110) crystals}
\author{Alexander V. Pavlov\inst{1, 2}
        \mail{a.pavlov@physics.spbstu.ru}%
        \and Andrei V. Korol\inst{3, 4}
        \and Vadim K. Ivanov\inst{1}
        \and Andrey V. Solov'yov\inst{4}
        \thanks{On leave from Ioffe Physical-Technical Institute, St. Petersburg, Russia.}%
}                     % Do not remove

\institute{Peter the Great St. Petersburg Polytechnic University, 
            Polytechnicheskaya 29, 195251 St. Petersburg Russia
            \and Research Institute for Nuclear Problems, Belarusian State University, 
            Bobruiskaya 11, 220030 Minsk Belarus
            \and St. Petersburg State Maritime University, Leninsky ave. 101, 
            198262 St. Petersburg Russia
            \and MBN Research Center, Altenh\"{o}ferallee 3, 
            60438 Frankfurt am Main Germany}
\date{Received: date / Revised version: date}
% The correct dates will be entered by Springer
%
\abstract{
In this paper we present the results of a systematic numerical analysis 
of the channeling properties 
for electrons and positrons in oriented straight and 
periodically bent diamond(110) crystals. 
We analyse the dependence of intensity of radiation emitted 
%%channeling and undulator radiation intensity 
on the projectile energy as well as on the bending amplitude.
%of the periodically bent crystal.
%
The analysis presented is based on grounds of accurate numerical simulations 
of a channeling process.
The simulation parameters, such as the crystal orientation, thickness and
bending parameters of the crystals as well as the energy of the projectiles, 
were chosen to match those used in past and ongoing experiments.
The peculiarities which appear in the radiation spectra are attributed to 
the interplay of various radiation mechanisms.
The analysis performed can be used to predict and explain future experimental results.
\PACS{
      {61.85.+p}{Channeling phenomena}   \and
      {41.60.-m}{Radiation by moving charges}  \and
      {41.75.Ht}{Relativistic electron and positron beams} \and  
      {02.70.Uu}{Applications of Monte Carlo methods}   \and
      {07.85.Fv}{X- and $\gamma$-ray sources, mirrors, gratings, and detectors} 
    } % end of PACS codes
} %end of abstract

\authorrunning{A. V. Pavlov, \textit{et. al}}
\titlerunning{Channeling of electrons and positrons in straight 
                and periodically bent diamond(110) crystals}

\maketitle

\section{Introduction}

Creation of new light sources is an important part of scientific progress. 
Nowadays, laser systems are capable of emitting electromagnetic radiation from infrared to 
X-ray range~\cite{emma2010first, mcneil2010x}. 
However, creation of the devices able to emit photons with sub-angstrom wavelength 
is still challenging. 
Meanwhile, such light sources open a number of new possibilities 
for various scientific experiments and technological applications
such as various medical applications, photon  induced disposing of a nuclear waste 
and nuclear reaction
\cite{solov2017multiscale, ledingham2003applications, ledingham2002laser, hajima2008proposal}. 

One of the promising ways of generating sub-MeV - MeV photons relies 
on channeling of ultra-relativistic charg\-ed particles through oriented crystals.
The basic effect of channeling in a straight crystal is in an anomalously
large distance which a positively charged projectile penetrates moving along a crystallographic
direction trapped in a potential well created by atomic planes or axes~\cite{lindhard1965influence}.
The particles which are trapped inside the potential well oscillate in 
the transverse direction while propagating in a planar or axial channel. 
The channeling oscillations result in a specific type of radiation - 
the channeling radiation  (ChR)~\cite{kumakhov1976theory}.
Its intensity depends on the type of a crystal and the crystallographic direction, 
the type of a projectile and its energy~\cite{andersen1983channeling, bak1985channeling, 
uggerhoj2005interaction, chirkov2016channeling}.

Channeling also occurs in bent crystals, when 
the bending radius exceeds a critical value~\cite{tsyganov1976estimates}. 
The motion in a bent crystal consists of two main components: 
the channeling oscillations and a circular motion along 
the bent channel center line.
The latter gives rise to a synchrotron-type radiation \cite{jackson2012classical}.
The total spectrum of radiation emitted by an ultra-relativistic particle 
channeled in a bent crystal bears features of both the ChR and synchrotron radiation 
\cite{taratin1998particle, solov1996channeling, shen2018channeling}.     

Another type of the radiation appears when an ultra-relativistic particle channels in 
a periodically bent crystal (PBC) \cite{korol1998coherent, korol1999photon}.
Such a system, called a crystalline undulator (CU), 
the undulator-type radiation (CUR) appears due to the periodicity in 
the projectile's trajectory which follows the shape of periodic bending.
By changing the type of a projectile, the projectile energy $\E$, 
the crystal type and the bending parameters (such as the bending amplitude $a$, 
period $\lamu$ and profile \cite{book2014channeling}) 
it is possible to tune the intensity and frequency of CUR.
%
% For projectiles with energy in range $\varepsilon \simeq 10^{-1} - 10^1$ GeV the range 
% of possible PBC parameters are as follows bending amplitude $a \simeq 10^0 - 10^1$ \AA{},
% bending period $\lamu \simeq 10^0 - 10^2 \mu$m.
%
By using CU it is possible to achieve the peak brilliance of CUR up to 
$10^{25}$ photons/s\, mrad$^2$ mm$^2$\, 0.1\%BW for photons in the 
energy range $10^{-2} - 10^1$ MeV \cite{book2014channeling}, 
such values cannot be achieved in the conventional 
undulators based on the magnetic field \cite{yabashi2017next}.
A systematic study of channeling and radiation parameters for various parameters 
of CU can open new opportunities for creation of new radiation sources. 

In recent years, the study of the channeling phenomenon has attracted much of attention. 
A significant number of theoretical 
\cite{shen2018channeling, kostyuk2013crystalline, korol2017channeling, 
      korol2016simulation, sushko2015multi, polozkov2014radiation, sushko2013silica, 
      sushko2013sub, pavlov2019interplay, agapev2018chan} 
and experimental
\cite{backe2018channeling, wistisen2017radiation, wistisen2016channeling, wistisen2014experimental,
      uggerhoj2015intense, wienands2015observation, bandiera2015investigation, backe2015channeling,
      mazzolari2014steering, bagli2014experimental}
works has been done to define the channeling parameters and 
electromagnetic radiation spectrum arising in the process.
Additionally, several experiments have  been performed aiming at detecting the CUR. 
The recent attempts include experiments with 195–855 MeV electron beam at the
Mainz Microtron (MAMI) facility \cite{p58_2016badems, backe2018channeling}
carried out with strained Si$_{1-x}$Ge$_x$ superlattices \cite{mikkelsen2000crystalline,
krause2002photon}, experiments at SLAC with $10 - 35$ GeV 
electrons and positrons of FACET beam \cite{wienands2016talk, wienands2017channeling} 
and experiments with 
few-GeV positrons at CERN with PBC based on boron doped diamond \cite{p38_2016badems}.
Unfortunately, these attempts have not been entirely conclusive \cite{tran2017synchrotron}.

In order to make experimental studies more focused, 
additional theoretical analysis is required.
In this work, we predict the evolution of the channeling properties and the radiation spectra for 
diamond(110) based CU \cite{backe2018channeling}.  
Drastic changes in the radiation spectra with variation of the bending amplitude $a$ 
of PBC are observed for different projectile energies $\E$.
The changes are sensitive to the projectile's charge.
% The spectra changes differently for positively (positrons) and (electrons) charged 
% particles. 
%
The analysis presented predicts the results of 
experimental observations and allows us to design experiment in a way to 
get desired photon spectrum basing on the calculations.

%
% Experiments in that field requires precisely grown CU and sources of particles with 
% energy in sub-Gev, GeV range and because of that they tend to be expensive. 
%

%
% Such calculations can significantly reduce costs of channeling experiments.

%
% One of the possible experiment setup in Europe seems to be Beam Test Facility, but 
% some improvements is needed \cite{backe2011future}.
% 

% 
% Thus, it becomes possible get desired photon spectrum basing on calculations.

\section{Methodology and simulation parameters}

In order to accurately predict experimental observations, one must rely on a software 
package which allows the accurate simulation of projectile motion in a crystalline 
environment. 
Because of that, \textsc{MBN Explorer} software package \cite{solov2012mesobionano} was used 
to model the motion of ultra-relati\-vi\-stic projectiles through the crystalline medium 
along with the dynamic simulation of crystalline structures in a course of motion 
\cite{sushko2013simulation}.
The computational framework of simulations is described in detail in Refs.
\cite{book2014channeling, sushko2013simulation}. 
The computations account for the interaction of projectiles with separate
atoms of the environment. 
%
% Because of variety of implemented inter-atomic potentials, the \textsc{MBN Explorer}
% supports rigorous simulations of various environments, including crystalline, 
% amorphous, or even biological ones. 
%
The simulated trajectories were used to compute 
the spectra of electromagnetic radiation.
%
%Because of that, \textsc{MBN Explorer}  software 
% package was used to simulate the crystalline structure and the trajectories of 
% ultra-relativistic particles inside that structure as well as to calculate spectral 
% and angular distributions of emitted radiation \cite{sushko2013simulation}. 
%
This computational approach has been benchmarked previously in 
Refs.\cite{sushko2013simulation, korol2016simulation, backe2018electron, book2014channeling}.

In this work, the calculations were performed for electrons and positrons 
propagating in the oriented diamond (110) crystal.
Interaction between the ultra-relativistic projectiles and the carbon atoms was 
simulated using the Moli\'{e}re potential \cite{moliere1947theorie}. 
The lattice temperature was set to 300K.
The crystal and the particle's beam parameters were chosen to match 
the experimental conditions at the MAMI facility 
\cite{backe2015channeling, backe2018channeling}. 
Namely, the energy $\E$ of the projectiles was considered within the 
range $270 - 855$ MeV.
% , the minimum energy follows from the applicability of the classical 
% relativistic molecular dynamics for the calculations of particle trajectories, meanwhile,
% the maximal energy is determined by the capabilities of the MAMI experimental setup.  
%
The diamond crystal thickness $\Lcr$ was set to 20 $\mu$m. 
The periodic bending was assumed to have a harmonic shape 
$S(z) = a \cos(2\pi z/ \lamu)$ where the coordinate $z$ is measured along the 
incident beam direction, $\lamu$ is the bending period which was set to 5 $\mu$m.
Examples of the systems calculated in similar geometries can be found in Ref. 
\cite{korol2017channeling}.  
% The incident beam direction was chosen to be $\langle 10, -10, 1 \rangle$
% in order to avoid an axial channeling.
%
All calculations were performed for the beams of particles with zero emittance.
The particle trajectories and electromagnetic spectra were calculated for the
straight crystal ($a$ = 0 \AA) and for PBCs with following bending amplitudes 
$a$ = 1.2, 2.5 and 4.0 \AA{}. 
For each simulation condition $N = 6000$ trajectories were calculated. 
Statistical uncertainties due to finite numbers of the simulated trajectories 
correspond to the 99.9~\% confidence interval 
(e.g. uncertainties which are shown by shaded area in the radiation spectra,
Figs. \ref{fig:ep_270_855_024}, \ref{fig:ep_270_855_4}, \ref{fig:trj_spectra_apex}(b),
\ref{fig:e-855-splitted}, were estimated as 
3.3~$\sigma$ with $\sigma$ being the standard deviation).

The radiation spectra of the particles were calculated according to the quasi-classical
formalism by Baier and Katkov (see Ref. \cite{katkov1998electromagnetic}).
The comparison between the fully quantum approach and Baier and Katkov formalism
\cite{wistisen2019complete} shows that the quasi-classical approach 
provides high calculation accuracy in application to the planar channeling problems.
The emission spectra %particles with energies $\E = 270$ and 855 MeV 
%and different bending amplitudes 
were carried out for the following two values of the detector opening angle:
$\theta_0$ = 0.24 $m$rad and $\theta_0$ = 4.0 $m$rad. 
These values correspond to the apertures with diameter 4 and 40 mm at MAMI. 
The opening angles can be compared to the natural emission angles 
$\gamma^{-1}$ ($\gamma = \E/\text{mc}^2$ stands for the Lorentz factor):
for $\E = 270$~MeV  $\gamma^{-1} \approx 0.2$ $m$rad and for $\E = 855$~MeV
$\gamma^{-1} \approx 0.6$ $m$rad.  
Thus, for both projectile energies the aperture with the opening angle
$\theta_{0} = 4$~$m$rad collects almost all emitted radiation.  

\section{Obtained results}

\subsection{Radiation properties}

One of the observables which can be measured in the channeling experiments 
with PBC is the spectrum of radiation emitted
by the beam of particles \cite{backe2018channeling}.
A particle, channeled in a PBC, undergoes two types of quasi-periodic motion:
the channeling oscillations and that due to periodicity of the bending.
These motions bear close resemblance with the undulating motion.
As a result, constructive interference of the waves emitted from the 
similar parts of the undulating trajectory are added together.
%
% % Spectral distribution in that case consists of a set of harmonics. 
%
For each value of the emission angle $\theta$ the spectral 
distribution consists of a set of narrow and equally spaced peaks (harmonics).
In the soft-photon limit, when the emitted energy $\hbar \om$ is small
compared to the projectile energy $\E$, the frequencies of $n$th harmonic $\om_{n}$ 
of ChR or CUR can be found from the relation
\begin{equation}
    \om_{n} = \frac{2 \gamma^2 \Om}{1 + \gamma^2 \theta^2 + K^{2} / 2} n, 
    \qquad n = 1, 2, 3, ...,
    \label{eq:omegas}
\end{equation}
where $\Om$ stands for the frequency of either channeling oscillations 
$\Om = \Om_{ch}$ or the CU oscillations, % one due to the bending period $\lamu$,   
$\Om = \Om_u = 2\pi/\lamu$.
$K^2$ is the mean square of the undulator parameter.
In a CU, the motion of a particle consists of two independent quasi-periodic modes, 
therefore, $K^2$ is given by a sum of squared undulator parameters corresponding to different modes: 
$K^2 = K^2_{\rm u} + K_{\rm ch}^2$,
where $K_{\rm u} = 2 \pi \gamma a / \lamu$ - the undulator parameter of a CU,
$K_{\rm ch}^2 = 2 \gamma^2  \langle v_{\perp}^2 \rangle / c^2$ - the undulator parameter 
related to the channeling motion, $\langle v_{\perp}^2 \rangle$ stands for the 
average velocity of the transverse motion
(see Ref.~\cite{book2014channeling} for details). 

To estimate the dependence of ChR and CUR spectral densities on the bending amplitude $a$, 
one notices that spectral density, $dE/ \hbar d \omega \equiv I$, of the radiation energy
emitted by a bunch of particles experiencing a quasi-periodic motion is proportional to 
\begin{itemize}
    \item average number $\meanN$ of particles participating in the motion,
    \item average distance $\Lch$ covered by a particle,
    \item squared Fourier image of the particle's acceleration, which can be written
            as $\Om^4 A^2$ with $\Om$ and $A$ standing for the frequency and the 
            (average) amplitude of the quasi-periodic motion.
\end{itemize}
As a result one arrives at:
\begin{equation}
    I \propto \meanN \Lch \Om^4 A^2\,.
    \label{eq:chan_int}
\end{equation}
% This general relation can be applied to the channeling and undulator motion 
% of electrons and positrons.
This estimate can be used to qualitativly explain the results of accurate calculations presented
further in the paper.

\begin{figure*}[]
    \centering
    \includegraphics[width=\textwidth]{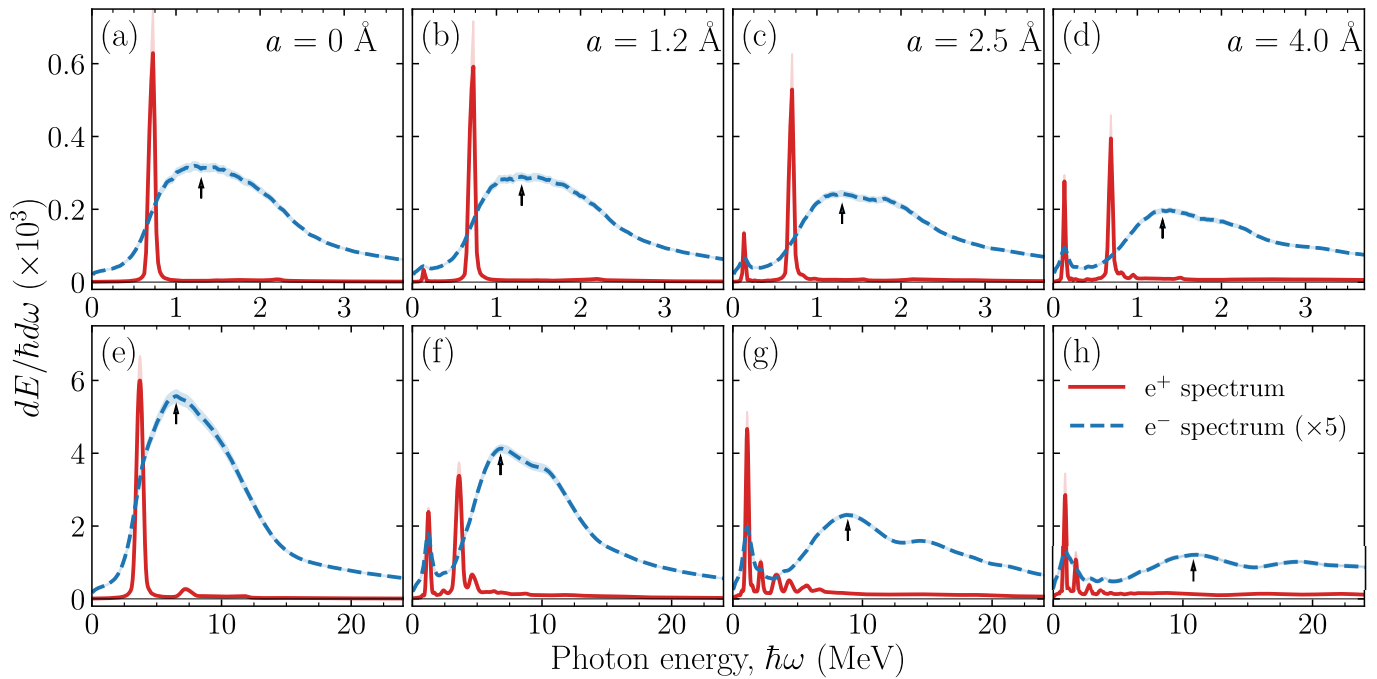}
    \caption{Spectral distributions of radiation by 270 MeV (upper row, graphs (a) - (d)) 
            and 855 MeV (lower row, graphs (e) - (h))
            electrons (dashed blue curves) and positrons (solid red curves)  calculated 
            for the opening angle $\theta_0 = 0.24$ $m$rad.
            First column, graphs (a) and (e), correspond to the straight crystal ($a = 0$ \AA{}).
%            The spectra for the straight crystal ($a$ = 0 \AA{}) are presented in the first column, graphs (a),(e).
            Other columns correspond to PBC with bending amplitudes 1.2, 2.5, 4.0 \AA{} as indicated
            in the upper graphs.
%             The spectra for PBC diamond (110) crystal with different bending amplitudes 1.2, 2.5, 4.0 \AA{} are
%             presented on the graphs (b), (f) and (c), (g) and (d), (h) respectively.
            Note that the electron spectra are multiplied by factor of 5. 
            Shading indicates the statistical error due to the finite number of 
            simulated trajectories. 
%             The upward arrows show the positions of ChR peak for electrons in straight and 
%             peak corresponding to volume reflection in PBCs 
%             (for detailed explanation see Section \ref{sec:overbarrier}).
            The upward arrows show the positions the peaks due to ChR (in straight crystal) and
            due to interplay of ChR and additional emission due to volume reflection in PBC 
            (see Section \ref{sec:overbarrier}).
            For the sake of comparison we quote the intensities of the background
            incoherent bremsstrahlung estimated within the Bethe-Heitler approximation:
            $2.9\times10^{-6}$ and $2.5\times10^{-5}$ for $\E=270$ and 855 MeV, respectively.
            }
            \label{fig:ep_270_855_024}
\end{figure*}

\begin{figure*}[]
    \centering
    \includegraphics[width=\textwidth]{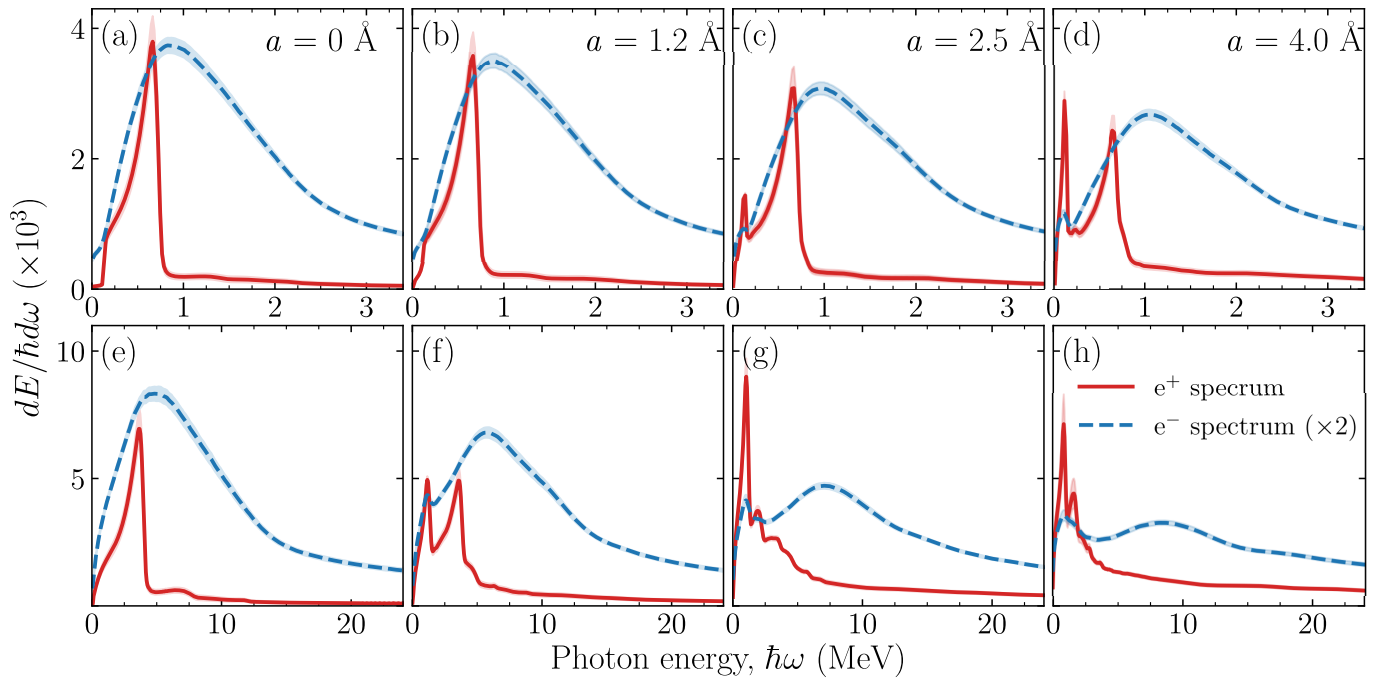}
    \caption{Same as in Fig. \ref{fig:ep_270_855_024} but for larger opening angle, $\theta_0 = 4 m$rad.
            The electron spectra are multiplied by 2. 
            The intensities of the background
            incoherent bremsstrahlung are
            $1.3\times10^{-4}$ and $1.8\times10^{-4}$ for $\E=270$ and 855 MeV, respectively.
%     The spectral distributions of radiation by 270 (a) - (d) and 855 (e) - (h)
%             MeV electrons and positrons and opening angle $\theta_0 = 4 m$rad.
%             The spectra for the straight crystal ($a$ = 0 \AA{}) are presented in the first column, graphs (a)(e).
%             The spectra for PB diamond (110) crystal with different bending amplitudes 1.2, 2.5, 4.0 \AA{} are
%             presented on graphs (b), (f) and (c), (g) and (d), (h) respectively.
%             For sake of comparison the electron spectra are multiplied by factor of 2. 
%             Shading indicates the statistical error due to the finite number of 
%             simulated trajectories. 
}
    \label{fig:ep_270_855_4}
\end{figure*}

The radiation spectra for the positrons and electrons with energies 
$\E = 270$ and $855$ MeV calculated for the opening angles $\theta_0 = 0.24$ and 
$4$ $m$rad are presented in Fig.~\ref{fig:ep_270_855_024} and 
Fig.~\ref{fig:ep_270_855_4}, respectively. 
%%%%%%%%%%%%%%%%%%%%%%%%%%%%%%%%%%%%%%%%%%%%%%%%%%%%%%

Figs. \ref{fig:ep_270_855_024}(a),(e) serve reference purposes 
and illustrate well-established features of the emission spectra in
straight crystals (see, e.g., Ref.~\cite{bak1985pedersen}). 
For both electrons ('-') and positrons ('+') the spectra are dominated by the peaks of ChR, 
the spectral intensity of which by far exceeds that of the bremsstrahlung 
radiation in the amorphous medium  (not shown in the figure).
For positrons, nearly perfect harmonic channeling oscillations
give rise to the narrow peak at $\hbar \om \approx 0.7$ MeV for $\E = 270$ MeV and 
at $\hbar \om \approx 3.6$ MeV for $\E = 855$ MeV.
Due to the strong anharmonicity of the electron channeling oscillations, the peaks of ChR (marked with the
upward arrows) are less pronounced and significantly broadened \cite{bak1985pedersen}
(note the scaling factor $\times5$ applied to the electron spectra). %in Fig.~\ref{fig:ep_270_855_024}).

%%%%%%%%%%%%%%%%%%%%%%%%%%%%%%%%%%%%%%%%%%%%%%%%%%%%%%%%%%%%%%%%%%%%%%%%%%% 

% % Let us now move from the analysis of well studied case of straight 
% % crystal to the case of PBC.
% 
To comment on the spectra formed in PBC we first consider the smaller energy of projectiles,
$\E = 270$ MeV. 
% % Consider first the emission spectra for projectiles with energy 
% % $\E = 270$ MeV 
% % (Figs.~\ref{fig:ep_270_855_024}(b -- d)). 
% 
% % In that case, the peak of CUR appears on the electron and positron emission 
% % spectra at the energy $\hbar \om \approx 0.13$ MeV.
In Figs.~\ref{fig:ep_270_855_024}(b -- d) the peak of CUR appears in both the electron and positron emission 
spectra at the energy $\hbar \om \approx 0.13$ MeV.
% 
% % With increase of the bending amplitude the intensity of CUR grows 
% % following Eq. \ref{eq:chan_int}.
The peak intensity increases with the bending amplitude in accordance with Eq. \ref{eq:chan_int}.
%
% % The ChR intensity, in contrast, slightly drops since $\meanN$ and $\Lch$ drops
% % with $a$ (see further Figs.~\ref{fig:e_270_855_length}(a -- b) and Table~\ref{tab:ch}).
On the contrary, the intensity of ChR exhibits moderate decrease as $a$ increases. 
% % slightly drops since $\meanN$ and $\Lch$ drops
% % with $a$ (see further Figs.~\ref{fig:e_270_855_length}(a -- b) and Table~\ref{tab:ch}).
%
These patterns are similar for electrons and positrons.

% % For $\E = 855$ MeV, on the other hand, the evolution of the radiation spectra 
% % is not so obvious as for the lower energy. 
For higher projectile energy, $\E = 855$ MeV, the evolution of the spectra 
is less straightforward.

The most prominent feature in the positron spectra is a strong suppression of the ChR peak
as $a$ increases.
Indeed, for $a=1.2$ \AA{}, Fig.~\ref{fig:ep_270_855_024}(f), 
the ChR peak intensity,  $I_{\text{ChR}}^{(+)}(a)$, 
drops by a factor of two compared to the straight crystal.
For larger bending amplitudes the ChR peak virtually disappears,
Figs.~\ref{fig:ep_270_855_024}(g -- h).
%
% This is not the case for the positrons with energy $\E = 270$ MeV. 

In case of electrons, the modifications of the spectra are much peculiar. 
The peak of CUR have maximum as a function of $a$ and become broader (more 
synchrotron like) for high bending amplitude (Fig. \ref{fig:ep_270_855_024} (h))
The intensity of ChR $I_{\text{ChR}}^{(-)}(a)$ for 
electrons does not fall off so dramatically as for positrons. 
The peak of ChR become more and more blueshifted with increase of $a$
(note the position of upward arrows in Fig.~\ref{fig:ep_270_855_024})
Additional peak appears in the high energy part of the radiation spectra
of PBCs (Figs. \ref{fig:ep_270_855_024} (f -- h)).

These changes are associated with the dynamics 
of both channeled and dechanneled particles.

Channeled particles mainly contribute to the CUR and ChR radiation.
The analysis of the behavior of channeled particles is presented 
in Section \ref{sec:electrons}). 
%
% For electrons the number of challenging/dechanneling acts on 
% crystal length become significant and dechanneled particles 
% become important for consideration.
%

Dechanneled particles in a PBC are mainly involved in two processes: 
volume reflection (VR) \cite{taratin1987volume, taratin1987deflection}
in the vicinity of points of maximum curvature and motion through parts of 
the crystal with small curvature and under small angles,
i.e. the over-barrier motion as it defined 
in continuous potential approximation.
These types of motion contribute to the different parts of 
radiation spectra.
The first lies in the energy domain of the ChR for a given bending amplitude.
The second is emitted at higher energies and reveals itself as additional 
peak on the radiation spectra.
The possibility of radiation emission by over-barrier particles in 
the field of PBC was discussed qualitatively using the theory of 
continuous potential approximation in Ref. \cite{shul2008new}. 
However, the analysis of such a phenomena using the accurate all-atom 
molecular dynamics provides more detailed information 
and predicts the effect not only qualitatively, but quantitatively.
The analysis presented in the Section \ref{sec:overbarrier} reveals 
the contribution from the channeling and dechanneled particles to 
the radiation spectra.

The spectra obtained at $\theta_0 = 4$ $m$rad bear close resemblance
to one collected at the angle $\theta_0 = 0.24$ $m$rad.
For example, the spectra in Figs.~\ref{fig:ep_270_855_4}(a) and \ref{fig:ep_270_855_4}(e) 
are dominated by the ChR.
The evolution of CUR with $a$ for electrons and positrons with $a$ have same pattern.
However, the left shoulders of the peaks are red-shifted because of the dependence of 
$\om_n$ on the collection angle (see Eq.~\ref{eq:omegas}).
This broadening makes some of the features less distinguishable.
For example, the additional high-energetic peak in the electron spectra 
is seen in the spectra for $\theta_0 = 0.24$ $m$rad (Figs. \ref{fig:ep_270_855_024} (b -- d)),
disappears for $\theta_0 = 4$ $m$rad (Figs. \ref{fig:ep_270_855_4} (b -- d)).
Additionally, larger opening angle leads to the larger spectral densities of radiation 
(note different scales in Figs.~\ref{fig:ep_270_855_024} and \ref{fig:ep_270_855_4}), 
the larger the opening angle, the more radiation enters the detector area
\cite{book2014channeling}.
Another interesting feature is that the radiation intensity for the positrons and 
electrons varies differently with changing of the aperture 
(scaling factor of the electron spectra $\times 5$ for collection angle $\theta_0 = 0.24$ 
$m$rad and factor $\times 2$ for $\theta_0 = 4$ $m$rad)

It is also worth noting, that in the positron emission spectra the higher harmonics 
occur in the spectra of CUR and ChR, the number and intensity of which changes 
with the energy of positrons.
The spectral distributions of ChR and CUR with several harmonics for the positrons 
are clearly distinguishable from the smooth curves of the electron spectra.
For $\E = 270$ MeV the additional harmonics of ChR are barely seen
(note small bump in Fig.~\ref{fig:ep_270_855_024}(a) around $\hbar \om \approx 2.1$ MeV), 
but for $\E = 855$ MeV the higher harmonics are clearly seen both for the
CUR (additional peaks in Fig. \ref{fig:ep_270_855_024}(e) around $\hbar \om \approx 7.5$ MeV) 
and the ChR (additional four equidistant peaks in Fig.~\ref{fig:ep_270_855_024}(h)).
This fact is in agreement with the theory of undulator radiation (see, e.g., 
\cite{book2014channeling, alferov1989radiation}). 
It predicts that when the undulator parameter 
$K \sim K_{\text{u}} = 2 \pi \gamma \frac{a}{\lamu} \lesssim 1$,
emission spectra must contain few harmonics, the intensities of which rapidly 
decrease with the harmonic number $n$. 
Higher harmonics are more pronounced for lower apertures 
(see Fig. \ref{fig:ep_270_855_024} (g) and Fig. \ref{fig:ep_270_855_4} (g)).

\subsection{Channelling properties}

Ultrarelativistic electrons and positrons propagate in crystals 
in different potentials \cite{book2014channeling}. 
Positrons move in the smooth harmonic potential, in contrast, 
electrons move in the strong\-ly anharmonic potential.
Obviously, the different potentials result to the different channeling properties. 

% Let us now analyze the channeling properties for electrons and positrons.
% %
% In the continuous potential approximation the stable channeling inside a crystal is possible
% when the centrifugal force is less than the maximum transverse force acting on the particle 
% in the planar potential \cite{tsyganov1976estimates, tsyganov1976some}. 
% % 
% In general, one can quantify this statement in terms of the dimensionless parameter
% (the bending parameter) 
% \begin{equation*}
%     C = \frac{\E}{R U'_{max}} < 1,
% \end{equation*}
% where $\E/R$ stands for centrifugal force for an ultrarelativistic particle, 
% $R$ is the curvature radius \cite{tsyganov1976estimates} and 
% $U'_{max} = 6.7$ GeV/cm the maximum derivative of interatomic potential 
% (transverse force) for the diamond at room temperature \cite{book2014channeling}.  

% Let us now define the parameters that describe the behavior of particles during 
% the propagation inside a crystal.
% % 

In the experiments, the particles can enter the crystal at any possible 
transverse coordinate.
So, in our simulations initial values of the transverse
coordinates of the particles must be randomized.
Because of this and thermal fluctuation of the crystalline environment 
not all projectiles start moving inside a crystal in the channeling mode. 
A commonly used parameter to quantify the latter property is the acceptance defined as the ratio
$\calA = \Nacc/N$ of the $\Nacc$ number of particles that start motion in the channeling regime 
to the total number of particles $N$. 
The non-accepted particles experience unrestricted over-barrier motion at the entrance but can
rechannel somewhere in the bulk.
For different theoretical approaches to the interaction of the projectiles 
with the crystalline environments, the different criteria distinguishing 
channeling and non-channeling motions of the particles can be introduced.
For example, in the continuous potential approximation the transverse (inter-planar) 
and the longitudinal motions of the projectiles are decoupled \cite{lindhard1965influence}. 
As a result, it is natural to define the channeling projectiles as those with 
the transverse energie not exceeding the height of inter-planar potential barrier. 
Within this framework, the acceptance $\calA$ is determined at the entrance of the
crystal and is defined as the ratio of the number of the particles with
the transverse energy less than the depth of potential well to the total number of the particles. 
In the simulations based on the solution of the equations of motion, 
a projectile interacts, as in the reality, with the individual atoms of the crystal
\cite{sushko2013simulation}.
The potential experienced by the projectiles varies rapidly in the course of their 
motion, that couples the transverse and the longitudinal degrees of freedom. 
Therefore, other criterion is required to select the channeling episodes of the projectile
motion, i.e. the acts of acceptance and re-channeling. 
In that case one can assume channeling to occur when a 
projectile, while moving in the same channel, changes the sign of transverse velocity at 
least two times \cite{sushko2013simulation}. 
Thus, the acceptance $\calA$ is determined not at the entrance but at some distance 
inside the crystal. 

Using above approach it is possible to introduce the characteristic lengths
of channeling motion. 
The average distance which an accepted particle passes inside a crystal 
before the dechanneling is called the penetration length $\Lp$.
The mean distance which particle pass inside a crystal in channeling mode is often called 
de-channeling length $\Lch$.
In relatively thick crystal multiple events of the channeling - dechanneling - re-channeling can occur.
So, to determine the total distance which particles propagate inside a crystal 
in the channeling regime one can introduce a total channeling length $\Ltot$. 
The total channeling length $\Ltot$ can be calculated by averaging the length of all channeling
segments over the total number of trajectories $N$.

The parameters which characterize electron and positron channeling 
in the crystals with different bending amplitudes $a$ are presented in Table~\ref{tab:ch}.

\begin{table*}[]
\caption{
    The acceptance $\calA$, penetration length $\Lp$ 
    and total channeling length $\Ltot$ for 
    electrons and positrons with energy $\E = 270$ and $855$ MeV in 
    the straight and periodically bent diamond(110) crystal.
    The first column indicates the bending amplitude $a$
    ($a = 0$ corresponds to the straight crystal).
    The second column shows the maximal bending parameter 
    $C$ for corresponding
    projectile energy and bending amplitude.
    Condition $C = 4 \pi^2 a \E \mathbin{/} \lamu^2 U'_{\text{max}} < 1$
    allows one to estimate whether or not channeling 
    can effectively occur 
    ($U'_{\text{max}} = 6.7$ GeV/cm the maximum gradient of 
    interatomic potential
    for diamond at room temperature \cite{book2014channeling}).
    The last column represents the peak intensity of CUR 
    $I_{\text{CUR}}^{(+)}$ $(\times 10^3)$ for positrons with certain
    energy.
    }
\label{tab:ch}
\begin{center}
\begin{tabular}{ ll lll lllc }
\hline
$a$ (\AA{}) & $C$ & $\cal A$ & $L_{p}$ ($\mu$m) & $L_{tot}$ ($\mu$m) 
                  & $\cal A$ & $L_{p}$ ($\mu$m) & $L_{tot}$ ($\mu$m) 
                  & $I_{\text{CUR}}^{(+)}$ $(\times 10^3)$ \\
\hline
\multicolumn{9}{c}{$\E = 270$ MeV} \\
\hline
\multicolumn{2}{c}{} & \multicolumn{3}{c}{Electrons $e^-$} 
                     & \multicolumn{4}{c}{Positrons $e^+$} \\ \cmidrule{3-9}
0   & 0     & 0.70 & 5.32 $\pm$ 0.21  & 8.85 $\pm$ 0.29 
            & 0.96 & 18.84 $\pm$ 0.18 & 18.42 $\pm$ 0.27 & 0.002 \\

1.2 & 0.07  & 0.61 & 4.90 $\pm$ 0.20  & 7.62 $\pm$ 0.25 
            & 0.93 & 18.53 $\pm$ 0.21 & 17.78 $\pm$ 0.32 & 0.034 \\

2.5 & 0.15  & 0.51 & 4.26 $\pm$ 0.19  & 5.96 $\pm$ 0.19 
            & 0.88 & 18.76 $\pm$ 0.19 & 17.06 $\pm$ 0.32 & 0.134 \\

4.0 & 0.24  & 0.40 & 3.72 $\pm$ 0.17  & 4.29 $\pm$ 0.13 
            & 0.76 & 18.16 $\pm$ 0.24 & 14.65 $\pm$ 0.38 & 0.277 \\
\hline
\multicolumn{9}{c}{$\E = 855$ MeV} \\
\hline
\multicolumn{2}{c}{} & \multicolumn{3}{c}{Electrons $e^-$} 
                     & \multicolumn{4}{c}{Positrons $e^+$} \\ \cmidrule{3-9}
0   & 0     & 0.72 & 10.85 $\pm$ 0.32 & 11.72 $\pm$ 0.32 
            & 0.96 & 19.24 $\pm$ 0.14 & 18.87 $\pm$ 0.21 & 0.03 \\

1.2 & 0.23  & 0.48 & 7.25 $\pm$ 0.33  & 7.89 $\pm$ 0.28  
            & 0.83 & 18.94 $\pm$ 0.17 & 16.71 $\pm$ 0.40 & 2.38 \\

2.5 & 0.48  & 0.30 & 4.98 $\pm$ 0.28  & 4.46 $\pm$ 0.14  
            & 0.60 & 16.77 $\pm$ 0.30 & 11.96 $\pm$ 0.48 & 4.67 \\

4.0 & 0.77  & 0.21 & 3.51 $\pm$ 0.17  & 2.42 $\pm$ 0.06 
            & 0.26 & 14.55 $\pm$ 0.61 & 5.92 $\pm$ 0.34 & 2.83 \\
\hline
\end{tabular}
\end{center}
\end{table*}

The acceptance $\calA$ and the total channeling length $\Ltot$ are maximal in
straight crystal and than gradually decrease with bending amplitude $a$.
This can be explained using a simple but illustrative model of continuous potential: 
effective potential drops with increase 
of bending amplitude due to the centrifugal force \cite{bak1985pedersen}.
The acceptances are relatively same in the straight crystal for $\varepsilon = 270$ 
and 855 MeV, but drop with $a$ significantly faster for $\varepsilon = 855$ MeV, 
because of the higher growth rate of the centrifugal force. 
The acceptance is larger for positrons than for electrons.
This fact can be explained in a following way: in an oriented crystal a projectile electron, 
being attracted to an atomic plane, gains the transverse energy  via the harder collisions 
with the crystal constituents and, thus, switches to the over-barrier motion much faster 
than a positron which experiences (on average) softer collisions.
Because of almost harmonic potential, positrons have much larger channeling length 
$\Lch$ compared to electrons and once a positron is accepted, it channels without dechanneling, 
but despite electrons the re-channeling events (capturing of the particle to the channel after 
free movement inside the crystal \cite{book2014channeling}) for positrons are rare.
The channeling length, however, growth linearly with a projectile energy 
(see Eq.~6.3 and 6.4 in Ref.~\cite{book2014channeling}), 
which leads to higher channeling lengths in straight crystal for projectiles with energy 855 MeV.
The evolution of the channeling lengths with $a$ is very similar to the  evolution of acceptance where 
increase in the centrifugal force leads to the strong suppression of the depth of potential well. 
For example for 270 MeV electrons $L_{tot}(a = 0\ \text{\AA}) / L_{tot}(a = 4.0\ \text{\AA}) \approx 2$ 
and $\approx 1.3$ for positrons, in contrast, for 855 MeV electrons the ratio $\approx 4.7$
and $\approx 3.2$ for positrons.

For more information about the channeling properties, parameters and visualization of
the channeling trajectories see for example Ref.~\cite{shen2018channeling} section 3.1 
and Ref.~\cite{book2014channeling} chapter 6.
Comparison between the experimental and theoretical data one can find, for example, 
in Ref.~\cite{backe2018channeling}.

\section{Discussion}

\subsection{Positron channeling}
\label{sec:positrons}

Let us now analyze the case of positron channeling.
In the planar channeling regime, a charged projectile moves along a planar 
direction experiencing a collective action of the electrostatic field of lattice
atoms \cite{lindhard1965influence}.
For a positively charged projectile, the atomic field is repulsive, so that the particle 
is steered into the inter-atomic region and oscillates (channels) in between two 
adjacent crystalline planes. 
At some stage, due to the collisions with the crystal constituents,
the transverse energy becomes large enough to allow particle to leave the channeling
mode, i.e. to dechannel.
The opposite process, the re-channeling, is associated with the capturing to the
channeling mode.
In a sufficiently thick crystal, a projectile can experience dechanneling and 
re-channeling several times in the course of propagation.
In this case, the quantity $\meanL$ in Eq.~\ref{eq:chan_int} 
should account for the length of all channeling
segments in the trajectory $\Ltot$.
However, for the crystals shorter than the dechanneling length the re-channeling events are rare
\cite{shen2018channeling, book2014channeling, korol2017channeling, korol2016simulation}.
The positrons with energy of several hundreds MeV have channeling length $\Lch \gg 100$ $\mu$m
\cite{biryukov2013crystal}, so in that case $\meanL$ can be calculated 
as the mean penetration $\Lp$ length (this value is depicted in column 7 of Table~\ref{tab:ch}).
Correspondingly, the value $\meanN$ can be associated with the number of accepted particles,
which is related to the acceptance $\calA$ (column 6 of Table~\ref{tab:ch}). 

These data allows one to qualitatively analyze the dependence of the 
intensity of the CUR $I_{\text{CUR}}^{(+)}$ on $a$ in case of positrons.
Using Eq. \ref{eq:chan_int} and taking into account that
the factor $\Om_{\rm u}^2$ is independent on the amplitude, one writes:
$I_{\text{CUR}}^{(+)}(a) \propto \calA(a) \Lp(a) a^2$.
Here, the term $\calA(a) \Lp(a)$ is a decreasing function of the bending amplitude, 
in combination with $a^2$, resulting function $I_{\text{CUR}}^{(+)}(a)$ 
have the maximum at some value of the bending amplitude
(see the last column of Table~\ref{tab:ch}).
The maximum at $a$ = 2.5 \AA{} is clearly seen for $\varepsilon = 855$ MeV, 
but for $\varepsilon = 270$ MeV the dependence 
$I_{\text{CUR}}^{(+)}(a)$ constantly growth for all observed $a$, so the maximum
should appear at large bending amplitudes than those studied in this work. 

\begin{figure*}
    \centering
    \includegraphics[width=\textwidth]{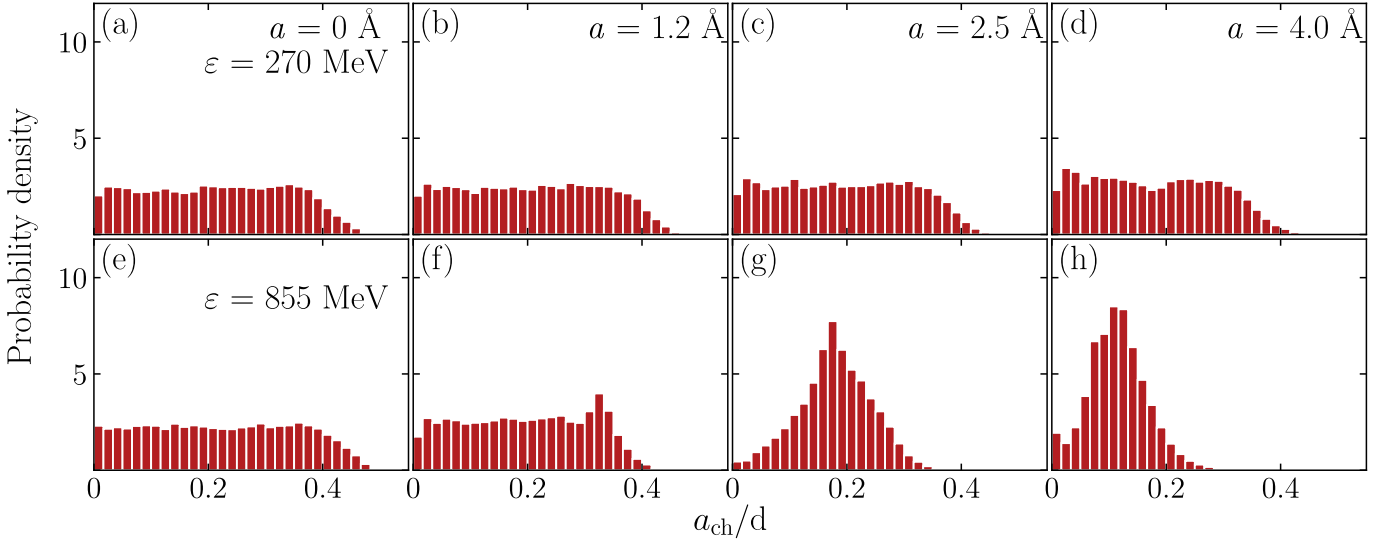}
    \caption{Probability density distributions over channeling amplitudes 
            $\ach$ for positrons with energies 270 and 855 MeV. 
            (a) - corresponds to case of the straight crystal case (bending amplitude $a$ = 0) 
            and energy $\E$ = 270 MeV, 
            (b)-(d) - corresponds to channeling in PBC with bending amplitudes $a$ = 
            1.2, 2.5 and 4.0 \AA{} correspondingly,
            (e)--(h) same as (a)--(d) but for positrons with energy $\E$ = 855 MeV}
    \label{fig:p_hyst}
\end{figure*}

Similar methodology is applicable to the analysis of dependence of ChR intensity 
$I_{\text{ChR}}^{(+)}$ on $a$. 
In this case, assuming the harmonic character of channeling oscillations, 
i.e. the independence of frequency $\Om_{\rm ch}$ 
on the amplitude of challenging oscillations 
$\ach$, one writes Eq. \ref{eq:chan_int} as follows:
$I_{\text{ChR}}^{(+)} \propto \calA(a) \Lp(a) \langle a_{\rm ch}^2\rangle(a)$.
Here, $\langle a_{\rm ch}^2\rangle(a)$ stands for the mean square amplitude
of channeling oscillations.
This quantity is a decreasing function of bending amplitude.
Indeed, as $a$ increases the centrifugal force, especially in the vicinity 
of the points of maximum curvature, drives the projectiles,
oscillating with large amplitudes, away from the channel.
This results in the strong quenching of channeling oscillations.

% Fig. \ref{fig:p_hyst} illustrates the distributions of number of particles
% on channeling oscillations amplitudes $\ach$ for different energies and bending amplitudes.
% %
% The amplitude is measured in units of $d$ and the maximum value equals to $\ach / d = 0.5$.
% The maximum amplitude of channeling oscillations equals to the half of interplanar distance 
% $d/2$ (d = 1.26 \AA{} for diamond(110)), probability distribution function are plotted 
% according to , so the maximum value on x-axis is 0.5.
%
Fig. \ref{fig:p_hyst} shows the probability densities for a particle to have a certain 
channeling amplitude, which is important for the analysis of the emission spectra 
of these particles.
These dependencies do not show the average deviation of the particle from the center of the channel.
Instead, they are defined as $(\Delta N (a) / \Delta a)/ N_{\text{tot}}$. 
Here $\Delta N(a)$ is the number of oscillations with the amplitude within the interval 
$\Delta a$, $N_{\text{tot}}$ – is the total number of oscillations (i.e. with any $a$).
It should be noted that the amplitude of oscillations is defined as the maximum deviation
of the channeling particle from the center of equilibrium.
Additionally, the fact that the center of potential shifts due to the centrifugal 
force (see Eq. A.25 in \cite{book2014channeling}) was taken into account.
% Because of centrifugal force the center of potential shifts.
% %
% For positrons the shift can be written in following form:
% $\rho(z) = S(z) - a/(\sigma^2 - 1) \cos(2\pi z / \lambda_u)$, where $S(z)$ stands 
% for the bending profile, second term for the centrifugal force and $\sigma^2 = 2 a / C d$.
%
Thus, the distributions were obtained as follows.
For each channeling particle, the distance from the equilibrium line were measured at 
the moment when the particle changes the sign of its velocity. 
After this, this procedure was repeated for all channeling sections of all particles. 
Based on the obtained values, the presented distributions are constructed. 
In this way, the presented distributions are obtained for channeling events throughout 
the depth of the crystal.

In case of a straight crystal, the particles are distributed 
uniformly over all possible oscillation amplitudes. 
The nature of this phenomenon stems from the quasi-harmonicity of the 
interatomic potential for positrons.
However, with increase of the bending amplitude, changes in the distribution occur.
For the positrons with energy $\E = 270$~MeV (Figs. \ref{fig:p_hyst}(a -- d))
the increase of $a$ leads to the small decrease in the maximum possible amplitude
and small reorganization of entire distribution.

For the positrons with energy $\E = 855$ MeV (Figs. \ref{fig:p_hyst}(e -- h))
the values of centrifugal forces are larger than for $\E = 270$ MeV, 
so distributions posses much more complicated evolution.
The probability density distribution for the PBC with $a = 1.2$ \AA{} looks similar to 
the distributions for $\E = 270$ MeV. 
Slight decrease in the channeling amplitude results 
in the small reorganization of the distribution.
In case of $a = 2.5$ and $4.0$ \AA{} (Figs.~\ref{fig:p_hyst}(g, h)) 
distributions have pronounced peak.
This behavior can be explained in a way that the centrifugal force moves potential 
minima closer to the point of maximum curvature, another words, the center of 
potential starts to shifts inside the channel.
In this situation positrons cannot precisely follow the minima of potential, 
so constant shift in the positron trajectories appears.
The peak in Figs.~\ref{fig:p_hyst}(g,~h) moves to lower values with increase of $a$, 
because particles with higher channeling amplitudes have higher probability 
to be kick-out from the channel.     

Summarizing above results let us conclude that the gradual decrease in all three factors, 
$\calA$, $\Lp$ and $\langle a_{\rm ch}^2\rangle$ results in the considerable drop 
in the intensity of ChR: the value of $I_{\text{ChR}}^{(+)}$ at $a=4$ \AA{} is 
by a factor of 30 less than for the straight channel, which is in accordance with the 
trend seen in Fig.~\ref{fig:ep_270_855_024} and \ref{fig:ep_270_855_4}.

%%%%%%FFT%%%%%%%%%%%
Another approach that can explain the behavior of ChR spectra is based on the analysis 
of fast Fourier (FFT) spectra of the particles trajectories.
This approach allows one to visualize the impact of bending amplitude
on the amplitude of channeling oscillations $\Om_{\rm ch}$.
On such dependence amplitude of oscillations $\ach \propto $ FFT intensity and
frequencies $\Om_{\rm ch} \propto $ Fourier frequency. 

\begin{figure*}
    \centering
    \includegraphics[width=\textwidth]{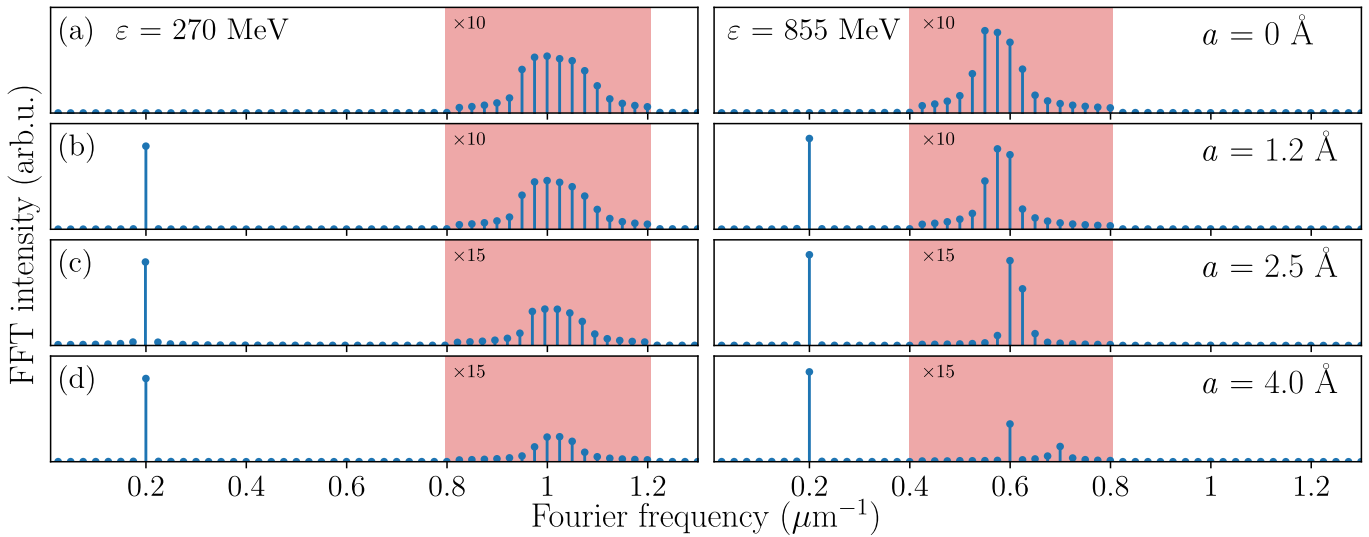}
    
    \caption{The FFT spectra of trajectories for positrons with energies 270 (left) and 
            855 MeV (right) which is channeled through the whole $L=20$ $\mu$m thick diamond(110).
            Graphs (a)-(d) refer to the straight crystal ($a=0$) and PBC with different $a =1.2, 2.5$
            and $4.0$ \AA{}, correspondingly.
            The red area shows the part of the spectrum due to the channeling oscillation 
            (note the scaling factor 10 for (a), (b) and 15 for (c), (d)). }
    \label{fig:p-fft}
\end{figure*}

Examples of FFT spectra are presented in Fig. \ref{fig:p-fft}.
The presented spectra are averaged over all particle that are moved through the crystal
without dechanneling. 
Fig. \ref{fig:p-fft} organized as follows: the FFT spectra are for positrons with 
energy $\E = 270$ MeV are shown on the left, spectra for $\E = 855$ MeV positrons 
are shown on the right.
Fig. \ref{fig:p-fft} (a) shows the FFT spectra for the straight crystal, in this case spectra 
are consist of peaks (the red shading) which corresponds to the channeling oscillations.
For the positrons with energy $\E = 270$ MeV oscillation frequencies lies between 
$f = 0.8 - 1.2$ $\mu\text{m}^{-1}$, however for positrons with energy $\E = 855$ MeV 
oscillations have frequencies between $f = 0.4 - 0.8$ $\mu\text{m}^{-1}$.
Difference in the frequency can be calculated using simple model (see e.g. chapter 4.2 in 
Ref.~\cite{book2014channeling}) where $\Om_{\rm ch} \sim \sqrt{U'/\text{d} m \gamma}$,
$U'$ is the first derivative of the inteplanar potential. 
Using above analysis one can get ratio 
$\Om_{\rm ch}(270)/\Om_{\rm ch}(855) = \sqrt{\gamma(855)/\gamma(270)} \approx 1.8$.
This value fits the observed results.

In Figs. \ref{fig:p-fft}(b -- d), corresponding to the PBCs, the FFT signals at 
$f=0.2$ $\mu$m$^{-1}$ 
correspond to the motion along the cosine centerline with the period $\lamu=5$ $\mu$m.
These peaks are normalized via division by $a$.
By comparing the spectra within the interval indicated in the red shading
for the straight and PBCs one sees the modification of the distribution 
of channeling oscillations.
For two energies spectra modify in two ways: for lower energy with increase of $a$ 
the intensity of channeling oscillations drops almost uniformly for all frequencies, for 
higher energy intensity of channeling oscillations quenches.
In particular, the FFT spectra clearly indicate that with increase in the
bending amplitude the ChR intensity decreases significantly, literally, only the oscillations on
one frequency survives. 
So, the FFT spectra clearly indicate that with the increase in
bending amplitude the ChR intensity decreases significantly for $\E = 855$ MeV and
uniformly drops for $\E = 270$ MeV.

To conclude the discussion of positron case, let us remark once again the dependence 
of readiation spectra on the projectile energy.
%
% Lower energies lead to weakening of the centrifugal force $\E/R_{\max}$ which, 
% in turn, lessens the suppression of the ChR contribution to the spectrum in a PBC.
%
Figs. \ref{fig:ep_270_855_024} and \ref{fig:ep_270_855_4} 
show the emission spectra for two different $\E$ and set of $a$. 
With increase in $a$, the ChR peak for the $\E = 855$ MeV projectiles virtually disappears, 
whereas for lower $\E = 270$ MeV it is still well pronounced.
This effect is visible for both small $\theta_0 = 0.24$ $m$rad 
and large $\theta_0 = 4$ $m$rad opening
angles and depends only on the strength of centrifugal force.

\subsection{Electron channeling}
\label{sec:electrons}

\begin{figure*}
    \centering
    \includegraphics[width=\textwidth]{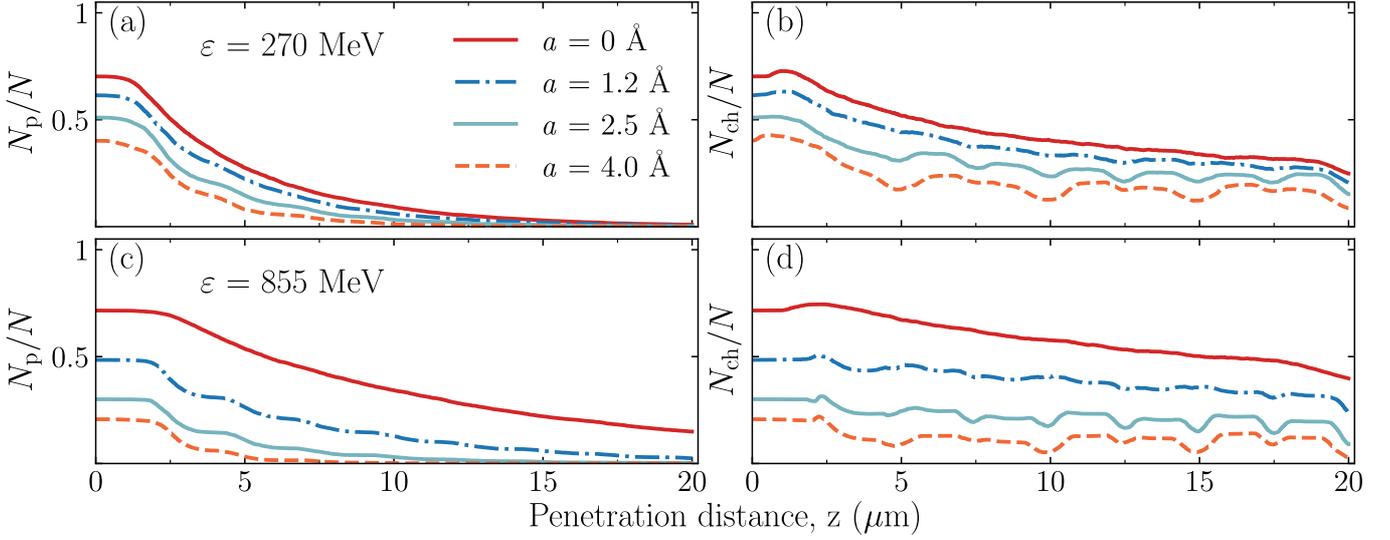}
    \caption{Fraction of channeling electrons in straight ($a$ = 0 \AA) and PBC diamond(110)
             crystal. 
             (a) primary fractions of the accepted particles for energy $\E$ = 270 MeV, 
             (b) fractions with account for the re-channeling for electrons with energy 
             $\E$ = 270 MeV,
             (c) - (d) same as (a) - (b) but for electrons with energy $\E$ = 855 MeV}
    \label{fig:e_270_855_length}
\end{figure*}

Let us turn back to the emission spectra of electrons in Figs.~\ref{fig:ep_270_855_024} 
and \ref{fig:ep_270_855_4}.
The dechanneling length $\Lch$ of electrons 
in the straight diamond(110) is less than the crystal thickness $L=20$ $\mu$m.
In case of PBC, $\Lch$ is even less being a decreasing function of $a$. 
To illustrate the differences in $\Lch$ for positrons and electrons as well 
as the dependencies on $a$, one can, firstly, compare the values of $\Lp$ for electrons 
(column 4) and positrons (column 7) in the Table~\ref{tab:ch}.
Main contribution to the low-energy part of the spectrum, where CUR dominates,
is made by the accepted particles.
To analyze the evolution of the CUR peak in the electron spectra,
let us start from Eq.~\ref{eq:chan_int} and then follow the arguments outlined 
in the positron case. 
Finally one can get following relation: 
$I_{\text{CUR}}^{(-)}(a) \propto \calA(a) \Lp(a) a^2$.
Similar to the positron case, the equation above is applicable when $\Lp(a)$ 
contains at least one CU period.

Analyzing the data from Table~\ref{tab:ch} one can notice that for the electrons
with energy $\E = 855$ MeV above statement is true only for $a=1.2$ and $2.5$ \AA{},
this explains the difference in the CUR intensities in such crystals 
(Figs.~\ref{fig:ep_270_855_024}(f,~g) and 
Figs.~\ref{fig:ep_270_855_4}(f,~g)). 
However, for PBC with $a = 4.0$ \AA{} the electrons with energy $\E = 855$ MeV 
have penetration depth less than one period $\Lp < \lamu$.
This leads to the fact that for PBC with $a=4$ \AA{}, the radiation becomes
more synchrotron-like. 
This manifests itself in the broadening of CU peak accompanied
by additional reduction in peak intensity 
(see Fig.~\ref{fig:ep_270_855_024}~(h) and Fig. \ref{fig:ep_270_855_4} (h)).

For the lower energy electrons $\Lp \approx \lamu$ for all bending
amplitudes.
To better illustrate this fact, let us look at the dependencies of the primary 
fraction of electrons on penetration distance.
The dependencies for $\E = 270$ MeV are shown in Fig.~\ref{fig:e_270_855_length} (a),
as one can notice for all bending amplitudes the primary fraction  
\text{virtually} disappears between 5 and 10 $\mu$m.
This means that for all bending amplitudes only a small fraction of 
electrons participate in the collective motion inside CU.
This fact leads to the conclusion, that the spectral density of CUR should increase with 
bending amplitude, since $a^2$ growth faster, than $\calA(a) \Lp(a)$ decreases.
For $\E = 855$ MeV (Fig.~\ref{fig:e_270_855_length}~(c)) the behavior is different,
the number of primary electrons on certain distance strongly depends on the bending amplitude.

Let us now analyze the part of spectrum where ChR dominates.
In this part of the spectrum the contribution of the re-channeled particles 
to the electron emission spectrum increases.

The interplay of several processes, occurring in the PBCs, leads to 
the structural transformations in this part of the spectrum.
This radiation is emitted by all particles experiencing the channeling motion.
These include the accepted electrons as well as those 
re-channeled anywhere inside the crystal.
It was shown in Ref. \cite{korol2017channeling} that periodicity in the bending 
significantly enhances the re-channeling rate even in the limit 
of large bending curvatures $1/R_{\max}$. 
At the distances with maximum curvature (the minima and maxima of the cosine bending profile), 
the dechanneling rate is the largest, this results to the minima 
of channeling fraction $N_{\text{ch}}/N$. 
In contrast, at the distances where the curvature approaches zero value, the re-channeling 
yields a significant increase in the number of channeling electrons. 
The dependence of $N_{\text{ch}}/N$ on $z$ allows one to calculate the 
total length $\Ltot$ of all channeling segments per a projectile.
The corresponding data are presented in the fifth column of 
Table~\ref{tab:ch}.

For the cosine bending profile, particles enter the PBC in the point
of maximum curvature. 
The centrifugal force filters the electrons with respect to their channeling 
amplitudes $\ach$, so similar to the positron case the \textit{accepted} 
particles oscillate with comparatively small amplitudes.

\begin{figure*}
    \centering
    \includegraphics[width=\textwidth]{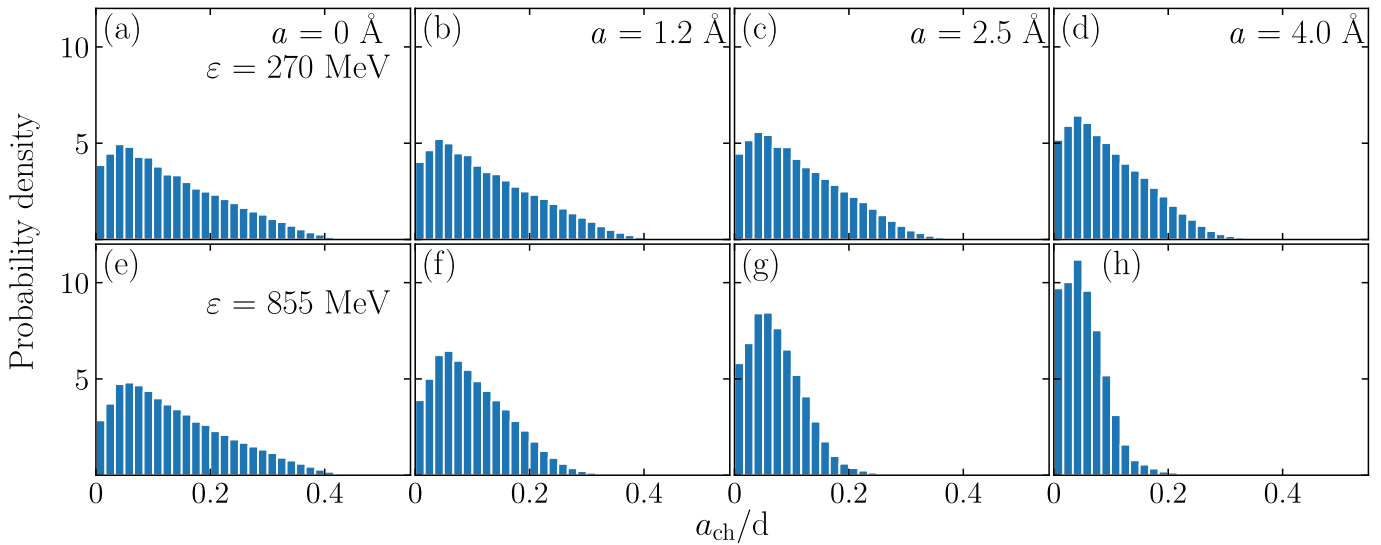}
    \caption{Probability density distribution over channeling amplitudes 
            $\ach$ for electrons with energies 270 and 855 MeV. 
            (a) - corresponds to straight crystal case (bending amplitude $a$ = 0) 
            and energy $\E$ = 270 MeV, 
            (b)-(d) - corresponds to channeling in PBC with bending amplitudes $a$ = 
            1.2, 2.5 and 4.0 \AA{} correspondingly,
            (e)--(h) same as (a)--(d) but for electrons with energy $\E$ = 855 MeV}
    \label{fig:e_hyst}
\end{figure*}

To illustrate this, the probability distributions of $\ach$ are shown 
in Fig.~\ref{fig:e_hyst}.
The distributions for electrons were obtained in the same way as distributions for positrons.
The distributions have asymmetric peak structure, where the lowest amplitudes are suppressed due 
to the scattering on a nuclear and the highest amplitudes are suppressed by the enhancement of 
de-channeling probability on the top of potential well.
The predominance of small amplitudes can be explained by the fact that the anharmonicity of 
channeling oscillation (see, e.g., Refs.~\cite{bak1985pedersen, korol2016simulation})
leads to the fact that low-amplitude oscillations occur with a higher frequency. 
As a result, the number $\Delta N(a)$ of small amplitude oscillations is larger than the number 
of oscillations with large amplitudes. 
Those, on average, they happen more often than the large amplitude oscillations. 
This is valid for both straight and bent crystals.
In Figs.~\ref{fig:e_hyst}(a -- d) the distributions for $\E = 270$ MeV are shown,
since for all observed bending parameters $C \ll 1$, 
distributions change slightly with the increase in the bending amplitude.
For $\E = 855$ MeV (Figs. \ref{fig:e_hyst}(e -- h)) changes are much more pronounced 
and increase in bending amplitude of PBC results in the significant reduction of 
$\ach$.
%
% Because of anharmonic character of the electron channeling oscillations, their frequency 
% depends on the channeling amplitude, being a monotonously decreasing function of latter 
% (see, e.g., Refs.~\cite{bak1985pedersen, korol2016simulation}). 
% %
% Therefore, the low-frequency channeling oscillations are suppressed.
% %
The emission frequency is related to $\Om_{\rm ch}$ as $\om_{\rm ch} \approx 2\gamma^2\Om_{\rm ch}$, 
one can conclude that the ChR spectrum of the accepted particles in PBCs is 
blue-shifted in comparison with the straight one, and this shift growth 
with the bending amplitude. 
The shift must be more pronounced for $\E = 855$ MeV than for $\E = 270$ MeV.
This feature is seen in Figs.~\ref{fig:ep_270_855_024}(a -- h) where
the upward arrows mark the positions of the ChR maximum.
The account of the emission from the re-channeled electrons allows one to explain why
the ChR spectrum does not decrease with $a$ so dramatically as it happens in the case of positrons.
Indeed, the re-channeling events occur on the (nearly) straight parts of the PB channel,
where the centrifugal force is small. 
Therefore, the channeling amplitude $\ach$ of these particles is uniformly distributed 
within the interval $\ach \lesssim d/2$ giving rise to the emission into the 
whole interval of the ChR energies.
As a result, the low-energy part of the ChR is non-zero for all amplitudes considered,
and the ratio of the maximum values of the intensity with good accuracy
follows the ratio of the $\Ltot$ values.

To conclude, the dechanneling--re-channeling dynamics together
with the strong anharmonicity of the channeling oscillations 
lead to the modifications in the shapes of the electron emission spectra. 
These changes are significantly different from those discussed for the positrons.

%%%%%%%%%%%%%%%%%%%%%%%%%%%%%%%%%%%%%%%%%%%%%%%%%%%%%%%%%%%%%%%%%%
\subsection{Radiation by dechanneled electrons}
\label{sec:overbarrier}

Let us now analyze the contribution to the emission spectra of electrons due to 
dechanneled particles.
%
% % The first manifestation of the importance of dechanneled
% % particles in a PBC is the additional peak in the emission spectra.  
Explicitly, this contribution reveals itself as an additional peak in the emission spectra from
a PBC.
%
% % This feature is more pronounced for 855 MeV electrons and 
% % appears in the spectra for $a=1.2$ \AA{} as a bump 
% % just beyond the powerful maximum of ChR (marked with upward arrows in Figs. \ref{fig:ep_270_855_024}) 
% % and it becomes more pronounced and shifts to higher energies as $a$ increases.
This feature is more pronounced for 855 MeV electrons and 
appears in the spectra for $a=1.2$ \AA{} as a bump 
just beyond the powerful maximum of ChR, see Fig. \ref{fig:ep_270_855_024}(b). 
It becomes more pronounced and shifted to higher energies as $a$ increases, Fig. \ref{fig:ep_270_855_024}(c, d).
%
% % In what follow the analysis of the role of different types of motions 
% % in formation of the radiation spectra is presented. 
In what follows we present the analysis of the role of different types of the motion 
experienced by channeling and non-channeling particles in formation of the emission spectrum. 
% We attribute this maxima to the radiation of the dechanneling particles.

%%%%%%%%%%%%%%%%%%%%%%%%%%%%%%%%%%%%%%%%%%%%%
\begin{figure*}[h]
    \centering
    \includegraphics[width=\textwidth]{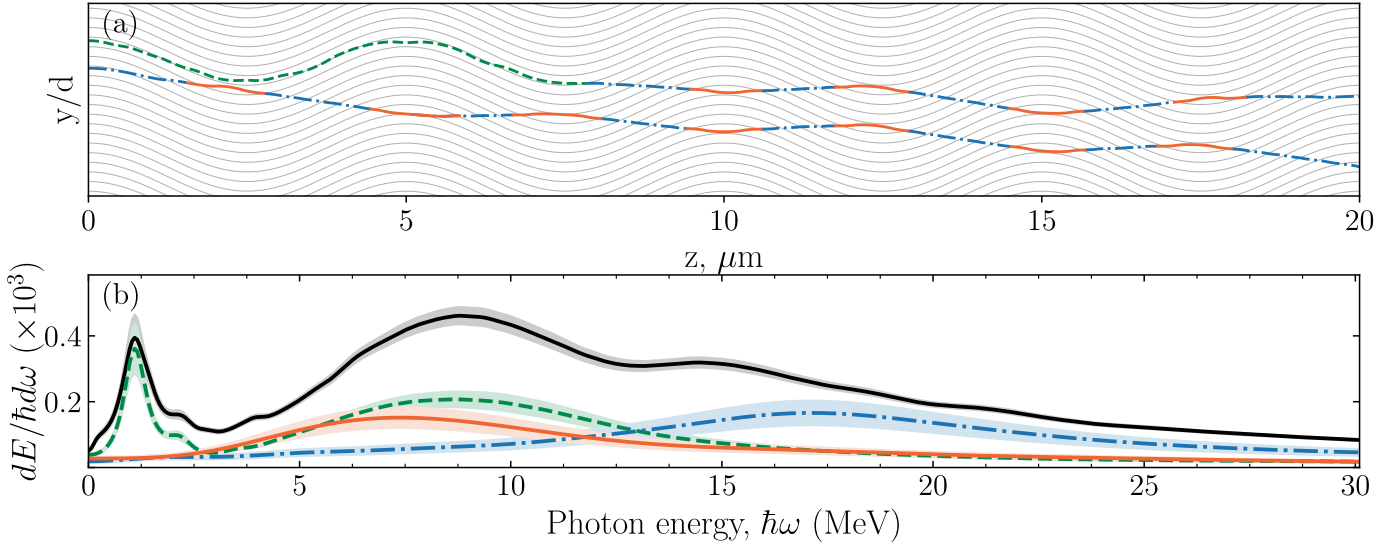}
%     \caption{(a) \textit{Exemplary} electron trajectories in a $a$ = 2.5 \AA{}.
%              The segments corresponding to  
%              challenging (dashed green lines), 
%              over-barrier (dotted-dashed blue line) and VR (solid orange line)
%              types of motion are highlighted.
%              Thin lines marks the atomic planes.
%              (b) Enhancement of the radiation over the Bethe–Heitler spectrum 
%              by 855 MeV electrons in PBC with $a = $ 2.5 \AA.
%              The color and line style of the spectra corresponds to the 
%              parts of the trajectories accordingly.
%              Solid black line indicates the total spectra.}
%              (b) Enhancement of the radiation emission over the Bethe–Heitler spectrum 
%               by a beam of 855 MeV electrons in the PBC.
%              Thick solid black curve indicates the total spectrum.
%              Dashed green, dashed-dotted blue and solid orange curves show the contribution 
%              from the segments of the channeling motion,
%              over-barrier motion and due to the VR, respectively.}
    \caption{(a) Selected simulated trajectories of 855 MeV electrons in a 
                PBC bent with $a$ = 2.5 \AA{}.
             Highlighted are the segments corresponding to the challenging regime (green dashed lines), 
             over-barrier motion (blue dashed-dotted line) and to VR events (solid orange line).
             Thin wavy lines mark the atomic planes.
             (b) Spectral distributions of radiation emitted by a beam of 855 MeV electrons in the PBC.
             Solid black curve indicates the total spectrum.
             Dashed green, dashed-dotted blue and solid orange curves show the contribution 
             from the segments of the channeling motion,
             over-barrier motion and due to the VR, respectively.
             The data refer to the opening angle $\theta_0=0.24$ $m$rad.
             }
    \label{fig:trj_spectra_apex}
\end{figure*}

% To illustrate the role of dechanneling particles, 
% the spectra from parts of the trajectories which correspond to 
% the channeling and dechanneling were calculated separately.
% 
% 
% To analyse the contributions to the radiation spectra from the
% channeling and non-channeling types of motion, 
% the radiation spectra from corresponding segments of 
% the trajectories should be calculated.
To compare the contributions to the total emission spectrum coming from 
channeling and non-channeling particles the following procedure has been adopted.
Each simulated trajectory has been divided into segments corresponding to different types of motion.
Namely, we distinguished the following parts of the trajectory:
(i) the channeling motion segments, 
(ii) segments corresponding to the 
over-barrier motion across the periodically bent crystallographic planes,
(iii) segments corresponding to the motion in the vicinity of points of maximum curvature 
where a projectile experiences volume reflection \cite{taratin1987volume,taratin1987deflection}.
For each type of the motion, the spectrum of emitted radiation has been computed 
as a sum of emission spectra from different segments.
Thus, the interference of radiation emitted from different segments has been lost.

The aforementioned procedure is illustrated by Fig. \ref{fig:trj_spectra_apex}.
Its upper panel, (a), presents two selected trajectories of 855 MeV electrons propagating in PBC with 
bending amplitude 2.5 \AA.
Different types of segments are highlighted in different colour and type of the line as 
indicated in the caption.
The emission spectra corresponding to different types of motion (calculated accounting for
all simulated trajectories) are shown in the lower panel (b).
The dependences presented allow one to associate the maxima in the total spectrum (black solid curve) 
with the corresponding type of motion. 
The radiation emitted from the segments of channeling motion (dashed green curve) 
govern the spectrum in the vicinity of the CUR peak ($\hbar\om_{\rm CUR}\approx 1$ MeV)
and provides a great contribution to the ChR, $\hbar\om_{\rm ch}\approx 6\dots12$ MeV.
Numerical analysis of the simulated trajectories has shown that the curvature of the trajectories
segments in the points of VR is close to that of the channeling trajectories.
As a result, the radiation from the VR segments is emitted in the same energy interval as ChR
so that the peak centered at $\approx 9$ MeV is due both to the channeling motion and to the 
VR events.
The over-barrier particles experience quasi-periodic modulation of the trajectory when crossing 
the periodically bent channels. 
The (average) period of these modulation is smaller than that of the channeling motion and
decreases with the increase of the bending amplitude.
For $a=2.5$ \AA{} this period is approximately two times less than the (average) 
period of channeling oscillations. 
As a result, radiation emitted from the over-barrier segments (dashed-dotted blue curve) 
is most intensive in the range $\hbar\om_{\rm ch}\approx 15\dots20$ MeV.
This contribution results in the additional structure in the total spectrum. 

\begin{figure}[h]
    \centering
    \includegraphics[width=0.48\textwidth]{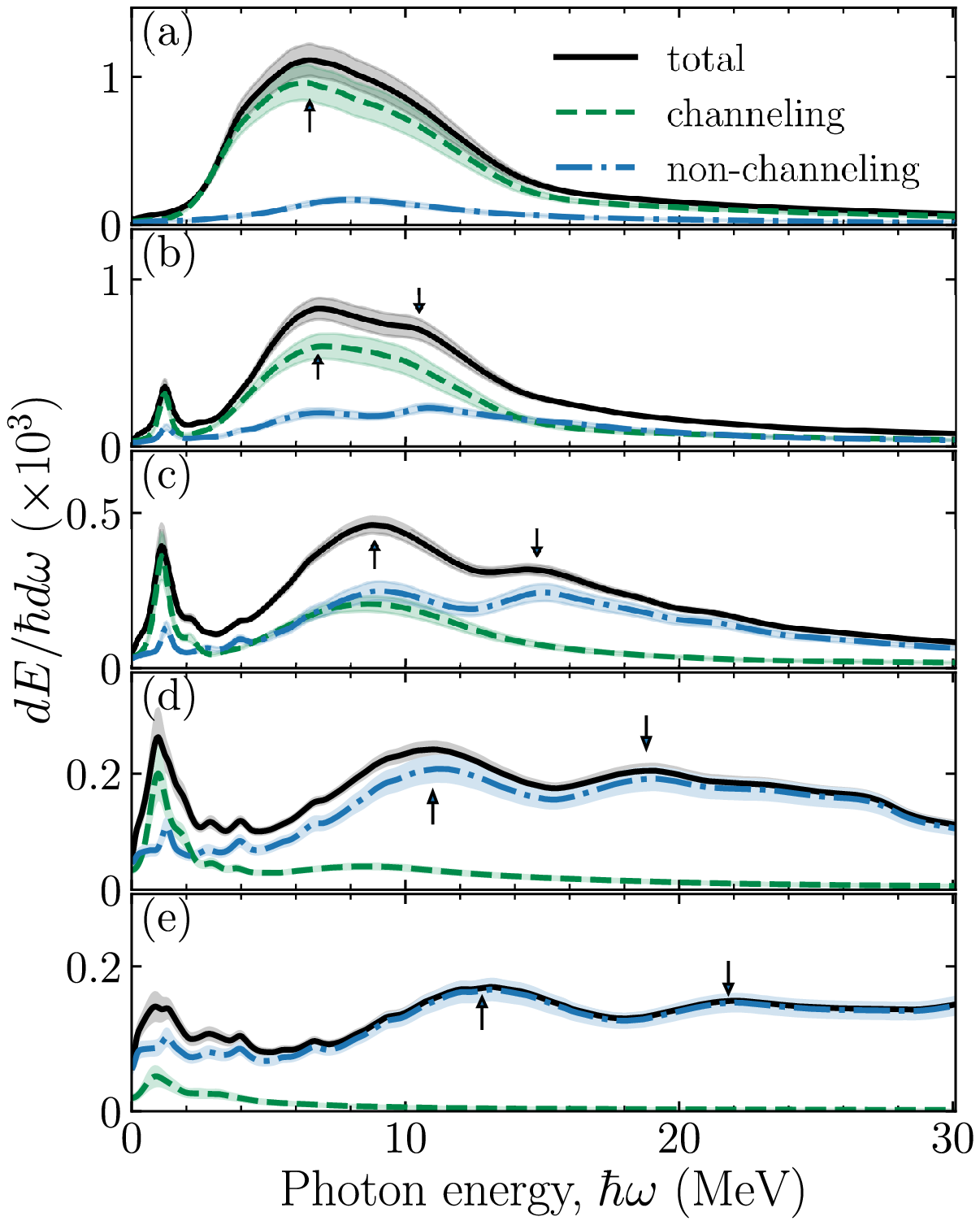}
    \caption{Spectral distributions of radiation emitted by 855 MeV electrons in straight, (a), and 
             periodically bent, (b)-(e), oriented diamond(110) crystal.
             Graphs (b)-(e) correspond to the bending amplitude $a = 1.2, 2.5, 4.0, 5.5$ 
             \AA, respectively.
             In each graph the solid black curve indicates the total spectrum, 
             the solid green curve shows the contribution from channeling segments,
             and dashed blue -- from all non-channeling segments (i.e., the over-barrier and the VR ones).
             The upward arrows indicate the maxima attributed to the ChR 
             (in the straight crystal) and to the interplay of ChR and VR 
             (in the PBC).
             The downward arrows show the position of the additional maxima
             appearing in the PBC due to the emission by over-barrier particles. 
             Shading indicates the statistical error due to the finite number of 
             simulated trajectories.
             The data refer to the opening angle $\theta_0=0.24$ $m$rad.}
% %     \caption{Enhancement factor of the radiation over the Bethe–Heitler spectrum
% %              by 855 MeV electrons in straight (a) and PBCs
% %              (b) -- (e).
% %              The bending amplitude of PBC was assumed to be equal $a$ = 1.2, 2.5, 4.0, 5.5 \AA.
% %              Solid black line indicates the total spectra, solid green the spectra 
% %              from channeling segments and dashed blue from dechanneling segments.
% %              The upward arrows indicate the maxima attributed to the ChR 
% %              in the straight crystal and the maxima due to volume reflection in 
% %              the PBC, the downward arrows show the position of the additional maxima
% %              appearing in the PBC. 
% %              Shading indicates the statistical error due to the finite number of 
% %              simulated trajectories.}
    \label{fig:e-855-splitted}
\end{figure}

To illustrate the the evolution of the contributions from the channelling and
non-channeling particles to the emission spectrum with bending amplitude.
we present Fig. \ref{fig:e-855-splitted}.
Graph (a) corresponds to the straight crystal,
graphs (b)-(e) -- to PBC with $a$ = 1.2, 2.5, 4.0 and 5.5 \AA, respectively.
Each graph presents the total spectrum, solid curve, 
the contribution of the channeling segments, dashed curve, 
and the contribution from the non-channeling segments (both over-barrier and VR), 
dash-dotted curve.

First, we discuss the modification of the spectra in the photon energy range 
far beyond the CUR peak.

In the straight crystal as well as in the PBC with small bending amplitude ($a=1.2$ \AA)
the emission spectrum is dominated by the channeling particles which provide main contributions
to the ChR peak.
As $a$ increases, the role of the non-channeling segments becomes more and more pronounced whereas
the channeling particles contribute less. 
The increase in $a$ leads to 
(i) increase of the curvature of a particle's trajectory in the vicinity of the VR points, 
(ii) decrease in the period of the quasi-periodic modulation of the trajectories of 
over-barrier particles.
As a result, two maxima seen in the graphs (b)-(e) become blue shifted as $a$ increases:
the maxima marked with upward arrows are due to the channeling motion and to the VR, 
those marked with downward arrows are associated with the over-barrier particles.
For large bending amplitudes, graphs (d)-(e), these maxima are virtually due to the emission
of the non-channeling particles only.

The low-energy part of the spectrum formed in PBC is dominated by the peak 
located at $\hbar\om\approx 1$ MeV. 
For moderate amplitudes, $a\leq 2.5$ \AA, when the bending parameter $C\ll 1$ (see Table~\ref{tab:ch}),
this peak associated with CUR and is due to the motion of the accepted particles which cover a distance 
of at least one period $\lamu$ in a periodically bent channel.
For larger amplitude, $a=4.0$ \AA{} ($C=0.77$), the penetration length $\Lp$ of the accepted particles 
become less than half a period leading to noticeable broadening of the CUR peak.
For even larger amplitudes, one notices further modifications of the peak which are related to 
the phenomenon different from the channeling in a periodically bent crystal.
Graph (e) shows the dependences for $a=5.5$ \AA{} which corresponds to the bending parameter larger than one,
$C=1.06$. 
As a result, only a small fraction of the incident electrons (less than 10 per cent) is accepted, 
and channels over the distance less than $\lamu/2$ having very small amplitude of channeling oscillations,
$a_{\rm ch}\ll d/2$.
Therefore, these particles virtually do not emit ChR but contribute to the CUR part of the spectrum
(see the dashed curve in the graph). 
However, this contribution is not a dominant one. 
The main part of the peak intensity in the total spectrum comes from the non-channeling particles,
see the dash-dotted curve.
The explanation is as follows. 
As discussed above, a trajectory of a non-channeling particle consists of short segments corresponding 
to VR separated by segments $\Delta z \approx \lamu/2$ where it moves in the over-barrier mode.
In the course of two sequential VR the particle experiences 'kicks' in the opposite directions,
see the lower trajectory in Fig. \ref{fig:e-855-splitted}(a)).
Therefore, the whole trajectory becomes modulated periodically with the period $2\Delta z \approx \lamu$.
This modulation gives rise to the emission in the same frequency as CUR. 

We believe, that the effects due to the interplay of different radiation mechanisms in PBC 
can be probed experimentally. 
In this connection we mention recent successful experiments on detection the excess of 
radiation in the emission spectra due to VR of 855 MeV electrons in 
oriented crystal bent with constant curvature \cite{mazzolari2014steering}.

\section{Conclusions}

In this work the channeling and radiation phenomena 
of 270 and 855~MeV electrons and positrons
in $\Lcr = 20\ \mu$m thick straight and periodically bent oriented diamond(110) crystals 
has been simulated by means of the
\textsc{MBN Explorer} software package \cite{solov2017multiscale, solov2012mesobionano}.
% % In this work the simulations of classical trajectories were performed for 
% % 270 and 855~MeV electrons and positrons
% % in $\Lch = 20\ \mu$m thick straight and periodically bent diamond(110) crystals 
% % by means of the channeling module \cite{sushko2013simulation} of the
% % \textsc{MBN Explorer} package \cite{solov2017multiscale, solov2012mesobionano}.
%

% % Electromagnetic radiation spectra were calculated from 
% % the simulated trajectories for electrons and positrons. 
%
Comparison of the emission spectra computed for two different opening angles, 
$\theta_0 = 0.24$ and 4 $m$rad,
has shown that although higher radiation intensity can be achieved for the larger opening angle, 
the use of the smaller aperture allows one to distinguish more peculiar features in the 
the spectra.

It has been demonstrated that the CUR spectral density for both positrons and electrons 
are non-monotonus functions of bending amplitude $a$ and projectile energy $\E$,
so that one can find the optimal parameters which maximize the yield of CUR. 
This result is important for the experimental studies of the radiation from CUs as well as for
designing and practical realization of periodically bent crystalline structures as the 
key element of the novel CU-based light sources.
%
% In case of positrons, for $\E = 270$ MeV CUR spectral density constantly growth with $a$,
% for $\varepsilon = 855$ MeV this dependence $I_{\text{CUR}}^{(+)}(a)$ have a maxima.
% %
% This opens new opportunities for creation of the CU-based light sources.
%
% Using the calculations one can explicitly select the parameters of periodically bent
% structure in order to maximize the yield of CUR or even create device with desired spectra.
%

Another important result of the analysis is that it is possible to vary the 
intensity of ChR by choosing the bending amplitude and projectile energy.
For positrons, ChR virtually disappears for the projectiles with energy 855~MeV
at much lower values of $a$ than for 270 MeV.
This provides an opportunity to decrease the intensity of ChR while keeping high yield of CUR.
This effect can be utilize to reduce the radiative energy losses
which are mainly due to ChR in oriented crystals . 
Low radiative energy losses maintain the stability of CUR \cite{korol2000total}.
The effects predicted in this paper can be measured in the experiments with sub-GeV
positron beams. 
Such beam is available at the DA$\rm \Phi$NE facility but some upgrades which will
improve the beam quality are desired \cite{backe2011future}.
% % on DA$\rm \Phi$NE electron–positron collider.
%

For electrons, it has been was shown that the intensity of CUR 
is much smoother function of bending amplitude.
It has been demonstrated that in the regime of large bending amplitudes the 
non-channeling particles provide significant contribution
resulting in additional structures in the emission spectra. 
%
% % The effects of volume reflection and over-barrier motion result 
% % in emission in two energy ranges.
%
These rather pronounced effects can be experimentally measured during 
using the high-quality electron beam available at the MAMI facility 
\cite{backe2018channeling, backe2015channeling, backe2018electron}.
%
% The experimental observations of above effects can be thought 
% as a fingerprint of high crystalline perfection of periodically bent structures.

\section{Acknowledgements}

The work was supported in part by the Alexander von Humboldt Foundation Linkage 
Grant and by the HORIZON 2020 RISE-PEARL project. 
We acknowledge the Supercomputing Center of 
Peter the Great Saint-Petersburg Polytechnic University 
(SPbPU) for providing the opportunities to 
carry out large-scale simulations.
We are grateful to Hartmut Backe and Werner Lauth (University of Mainz) for
useful discussions.
We are grateful to Referees for the constructive suggestions which lead to 
significant improvement of the manuscript.

\section*{Author contribution statement}
A.V.P. participated in setting up the parameters of CUs
to study and conducted a major part of simulations and
analysis of the raw data and was the main writer of the paper. 
A.V.K. participated in setting up the parameters of CUs
to study and analysis and discussion of the raw data and final results 
as well as in writing of the paper. 
V.K.I. participated in the analysis and discussion of the results.
A.V.S. led the studies and interpretation of the results.

\bibliographystyle{epj}
\bibliography{lib_chan}

\end{document}